\documentclass[aps,prl,twocolumn,superscriptaddress]{revtex4-1}
\usepackage{graphicx}
\usepackage{dcolumn}
\usepackage{bm}
\usepackage{amssymb}
\usepackage{mathrsfs}
\usepackage{amsmath}
\usepackage{comment}
\usepackage[caption=false]{subfig}
\usepackage{newtxtext}

\begin{document}
\preprint{preprint number}
\title{First Search for the $K_L \to \pi^0 \gamma$ Decay }

\newcommand{\InstKorea}{\affiliation{Department of Physics, Korea University, Seoul 02841, Republic of Korea}}
\newcommand{\InstOsaka}{\affiliation{Department of Physics, Osaka University, Toyonaka, Osaka 560-0043, Japan}}
\newcommand{\InstMichigan}{\affiliation{Department of Physics, University of Michigan, Ann Arbor, Michigan 48109, USA}}
\newcommand{\InstChicago}{\affiliation{Enrico Fermi Institute, University of Chicago, Chicago, Illinois 60637, USA}}
\newcommand{\InstNTU}{\affiliation{Department of Physics, National Taiwan University, Taipei, Taiwan 10617, Republic of China}}
\newcommand{\InstArizona}{\affiliation{Department of Physics, Arizona State University, Tempe, Arizona 85287, USA}}
\newcommand{\InstSaga}{\affiliation{Department of Physics, Saga University, Saga 840-8502, Japan}}
\newcommand{\InstKyoto}{\affiliation{Department of Physics, Kyoto University, Kyoto 606-8502, Japan}}
\newcommand{\InstKEK}{\affiliation{Institute of Particle and Nuclear Studies, High Energy Accelerator Research Organization (KEK), Tsukuba, Ibaraki 305-0801, Japan}}
\newcommand{\InstYamagata}{\affiliation{Department of Physics, Yamagata University, Yamagata 990-8560, Japan}}
\newcommand{\InstJeonbuk}{\affiliation{Division of Science Education, Jeonbuk National University, Jeonju 54896, Republic of Korea}}
\newcommand{\InstJPARC}{\affiliation{J-PARC Center, Tokai, Ibaraki 319-1195, Japan}}
\newcommand{\InstJINR}{\affiliation{Laboratory of Nuclear Problems, Joint Institute for Nuclear Researches, Dubna, Moscow region 141980, Russia}}
\newcommand{\InstNDA}{\affiliation{Department of Applied Physics, National Defense Academy, Kanagawa 239-8686, Japan}}
\newcommand{\InstOkayama}{\affiliation{Research Institute for Interdisciplinary Science, Okayama University, Okayama 700-8530, Japan}}

\author{N.~Shimizu}\InstOsaka
\author{J.~K.~Ahn}\InstKorea
\author{B.~Beckford}\InstMichigan
\author{M.~Campbell}\InstMichigan
\author{S.~H.~Chen}\InstNTU
\author{J.~M.~Choi} \InstKorea
\author{J.~Comfort}\InstArizona
\author{K.~Dona}\InstMichigan
\author{M.~S.~Farrington} \InstChicago
\author{N.~Hara}\InstOsaka
\author{H.~Haraguchi}\InstOsaka
\author{Y.~B.~Hsiung}\InstNTU
\author{M.~Hutcheson}\InstMichigan
\author{T.~Inagaki}\InstKEK
\author{M.~Isoe} \InstOsaka
\author{I.~Kamiji}\InstKyoto
\author{E.~J.~Kim}\InstJeonbuk
\author{J.~L.~Kim}\thanks{Present address: Jeonbuk National University, Jeonju 54896, Republic of Korea.}\InstKorea
\author{H.~M.~Kim}\InstJeonbuk
\author{T.~K.~Komatsubara}\InstKEK\InstJPARC
\author{K.~Kotera}\InstOsaka
\author{J.~W.~Lee}\InstKorea
\author{G.~Y.~Lim}\InstKEK\InstJPARC
\author{C.~Lin}\InstNTU
\author{Q.~S.~Lin}\InstChicago
\author{Y.~Luo}\InstChicago 
\author{T.~Mari}\InstOsaka
\author{T.~Matsumura}\InstNDA
\author{D.~Mcfarland}\InstArizona
\author{N.~A.~McNeal} \InstMichigan
\author{K.~Miyazaki}\InstOsaka
\author{R.~Murayama}\thanks{Present address: RIKEN Cluster for Pioneering Research, RIKEN, Wako, 351-0198, Japan.}\InstOsaka 
\author{K.~Nakagiri}\thanks{Present address: KEK, Tsukuba, Ibaraki 305-0801, Japan.}\InstKyoto
\author{H.~Nanjo}\InstOsaka
\author{H.~Nishimiya}\InstOsaka
\author{Y.~Noichi} \InstOsaka
\author{T.~Nomura}\InstKEK\InstJPARC
\author{T.~Nunes}\InstOsaka
\author{M.~Ohsugi}\InstOsaka
\author{H.~Okuno}\InstKEK
\author{J.~C.~Redeker} \InstChicago
\author{K.~Sato}\thanks{Present address: Institute for Space-Earth Environmental Research, Nagoya University, Nagoya, Aichi 464-8601, Japan.}\InstOsaka
\author{T.~Sato}\InstKEK
\author{Y.~Sato}\InstOsaka
\author{T.~Shimogawa}\thanks{Present address: KEK, Tsukuba, Ibaraki 305-0801, Japan.}\InstSaga
\author{T.~Shinkawa}\InstNDA
\author{S.~Shinohara}\thanks{Present address: Department of Physics, Osaka University, Toyonaka, Osaka 560-0043, Japan.}\InstKyoto 
\author{K.~Shiomi}\InstKEK\InstJPARC
\author{R.~Shiraishi} \InstOsaka
\author{S.~Su}\InstMichigan
\author{Y.~Sugiyama}\thanks{Present address: KEK, Tsukuba, Ibaraki 305-0801, Japan.}\InstOsaka
\author{S.~Suzuki}\InstSaga
\author{Y.~Tajima}\InstYamagata
\author{M.~Taylor}\InstMichigan
\author{M.~Tecchio}\InstMichigan
\author{M.~Togawa}\thanks{Present address: KEK, Tsukuba, Ibaraki 305-0801, Japan.}\InstOsaka
\author{T.~Toyoda} \InstOsaka
\author{Y.~C.~Tung}\thanks{Present address: National Taiwan University, Taipei, Taiwan.}\InstChicago
\author{Q.~H.~Vuong} \InstOsaka
\author{Y.~W.~Wah}\InstChicago
\author{H.~Watanabe}\InstKEK\InstJPARC
\author{T.~Yamanaka}\InstOsaka
\author{H.~Y.~Yoshida}\InstYamagata

\collaboration{KOTO Collaboration}\noaffiliation 

\date{\today}

\begin{abstract}
We report the first search for the $K_L \to \pi^0 \gamma$ decay, which is forbidden by
 Lorentz invariance, using the data
 from 2016 to 2018 at the J-PARC KOTO experiment.
 With a single event sensitivity of $(7.1\pm 0.3_{\rm stat.} \pm 1.6_{\rm syst.})\times 10^{-8}$,
 no candidate event was observed in the signal region.
 The upper limit on the branching
 fraction was set to be $1.7\times 10^{-7}$ at the 90\% confidence level.
\end{abstract}

\maketitle

The $K_L \to \pi^0 \gamma$ decay is forbidden by
 the conservation of angular momentum.
 In the $K_L$ rest frame, the spin of a massless photon must be polarized along the decay axis,
 but the back-to-back configuration of two-body decays
 does not allow the parallel component of the orbital angular momentum.
 In the broader context, $K_L \to \pi^0 \gamma$ threatens
 Lorentz invariance and gauge invariance~\cite{decayofheavymesons}.
 Such restrictions on $K_L \to \pi^0 \gamma$ provide the opportunity
 to search for new physics beyond the standard model (SM).
 In particular, as Ref.~\cite{HighEnergytest} suggests,
 similarly to experiments such as
 ones using optical resonators~\cite{lv_spect,lv_spect2},
 Lorentz invariance should be tested in short distances.
 Several scenarios predict a
 finite rate of the $K_L \to \pi^0 \gamma$ decay~\cite{decayofheavymesons,noncummutivity}.
 Using charged kaons, the E949 experiment at BNL searched for
 the $K^+ \to \pi^+ \gamma$ decay and
 set upper limit on the branching fraction to be
 $2.3 \times 10^{-9}$~\cite{e949results} at the 90\% confidence level (C.L.); 
 no measurements have been made for neutral kaons.

The KOTO experiment is being carried out using
 the 30~GeV Main Ring accelerator at
 J-PARC in Ibaraki, Japan. 
 A $K_L$ beam was produced by
 protons hitting a gold target, and was transported into the KOTO detector
 at an angle of $16^\circ$ from the primary beam~\cite{citebeamline}.
 Photons in the beam were removed by a 35-mm-thick lead plate placed in the upstream, and
 charged particles were removed by a sweeping magnet.
 The solid angle of the neutral beam after a collimation
 was 7.8~$\mu$sr, and the size was $8\times8~\mathrm{cm}^2$ at 20~m
 downstream from the target.  
 At the exit of the beam line, the peak of the $K_L$ momentum distribution was $1.4$~GeV$/c$.
 The $K_L$ incident rate to the KOTO detector was measured to be 7~MHz at a beam power of 50~kW,
 based on the measured $K_L \to 2\pi^0$ and $K_L \to 3\pi^0$ decays
 (this corresponded to $2\times 10^{-7}$ $K_L$ per proton on target).

The primary purpose of the KOTO experiment is to study the CP-violating
 $K_L\to \pi^0 \nu \bar{\nu}$ decay, which
 is suppressed in the SM, and the branching fraction is
 predicted to be $(3.0 \pm 0.3)\times 10^{-11}$~\cite{KLpi0nunutheory}.
 The signature of $K_L\to \pi^0 \nu \bar{\nu}$ is $2\gamma$ + {\it nothing};
 hence the KOTO detector consists of a fine-grained electromagnetic calorimeter
 and hermetic veto counters surrounding the decay volume.
 Thus, the apparatus is ideal to search for the $K_L\to \pi^0\gamma$ decay.

\begin{figure*}
\includegraphics[width=17cm]{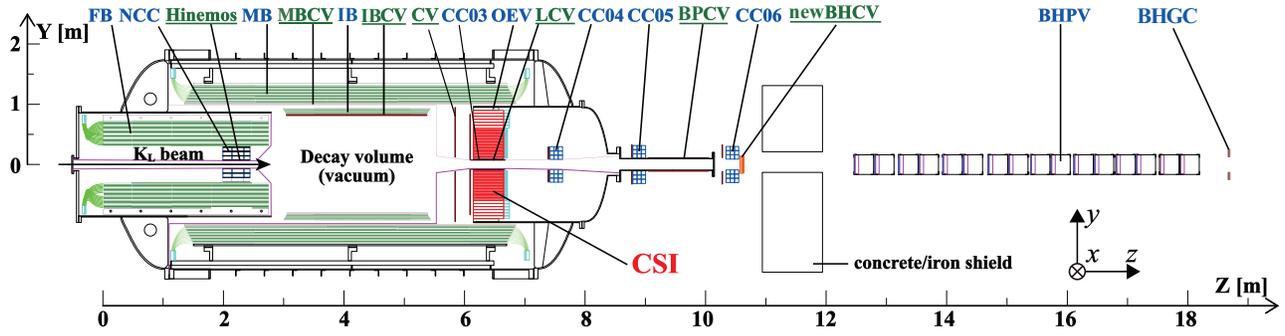}  
	\caption{Sectional view of the KOTO detector.
	The $K_L$ beam, pointing in the $+z$ direction in the figure,
	was transported from the left side.
	The names without (with) underline were
	for neutral (charged) particles.
	The origin of $z$ axis is the upstream edge of FB.}	\label{kotodet}
\end{figure*}

Figure~\ref{kotodet} shows the
 sectional view of the KOTO detector,
 in which the $z$ axis is in the center of the beam line.
 The decay volume was
 kept in vacuum of $10^{-5}$~Pa
 to suppress interactions between beam neutrons and residual gas.
 An electromagnetic calorimeter (CSI), consisting of 2716 50-cm-long undoped Cesium Iodide
 crystals, measured
 the energies and hit positions of incident photons from $K_L$ decays.
 The central and outer regions of CSI were made of $2.5\times 2.5~\mathrm{cm^2}$ and
 $5.0\times 5.0~\mathrm{cm^2}$ crystals in cross section,
 respectively~\cite{citeCSI, citeCSI2}.
 The crystals were stacked inside
 a 1.9-m-diameter cylinder, leaving a central
 $20\times 20~\mathrm{cm}^2$ hole for the beam path. 
 The FB, MB, and IB~\cite{IBpaper} were lead-scintillator sandwich counters,
 hermetically covering the decay volume to veto extra particles from $K_L$ decays.
 The inner surfaces of IB, MB, and the beam hole of CSI were covered with
 plastic scintillators
 (their thicknesses were 5~mm, 10~mm, and 3~mm, respectively)
 to veto charged particles.
 Two layers of
 3-mm-thick plastic scintillation counters (CV)~\cite{citeCV} were placed upstream
 of CSI to veto charged particles.
 To veto charged particles escaping through the beam hole of CSI, 
 three layers of wire chambers 
 were placed downstream of CSI in the beam (newBHCV).
 To veto photons passing through the beam hole,
 four sets of collar-shaped undoped Cesium Iodide
 counters were installed in CSI (CC03) and downstream of CSI (CC04, CC05, and CC06).
 To veto photons passing through the beam hole,
 sixteen modules, each made of lead and aerogel (BHPV)~\cite{cite_BHPV},
 and four modules, each made of lead and acrylic plates (BHGC),
 were placed downstream of CSI. 
 To veto particles going upstream,
 a counter made of undoped Cesium Iodide crystals (NCC) was
 placed inside of FB.
 The waveform of the signal from all the detector components
 was recorded with either 125~MHz or 500~MHz
 digitizers. 
 Details of the KOTO detector are available in Refs.~\cite{KOTOdet,KOTO2015}. 
 
In this analysis, we used data taken in the periods of
 from May to June 2016, from May to July 2017,
 and from January to February 2018
 with the proton beam power of 42-50~kW, corresponding to
 $2.8\times 10^{18}$ protons on target in total.
 The trigger required an energy deposit of $>550$~MeV in
 CSI with no coincident
 signals in IB, MB, CC03, CC04, CC05, CC06, CV, and NCC.
 The online energy thresholds for the
 veto counters were set
 sufficiently higher than those used in the offline analysis to avoid acceptance loss.
 With CSI, the number of clusters was calculated online. A cluster is
 defined as a collection of contiguous crystals with energies
 deposited larger than 22~MeV and 44~MeV for the small and large crystals, respectively.
 The data with exactly three clusters in CSI was used to search
 for the $K_L \to \pi^0\gamma$ decay. 
 Details of data acquisition system are available in Ref.~\cite{JayDAQpaper}. 
 
Candidates for $K_L \to \pi^0\gamma$ were required to have exactly
 three clusters, $\gamma_0, \gamma_1$, and $\gamma_2$
 (hereafter, $1, 2$ are indices of two $\gamma$'s from $\pi^0$ decay and
 0 is for the other $\gamma$) in CSI.
 Each cluster was reconstructed by integrating
 all adjoining crystals located within 70~mm and with the deposited energy larger than 3~MeV~\cite{cite_masuda}.
 The cluster energy was defined as the sum of all the energy deposits of
 the crystals in the cluster.
 The cluster timing and $xy$ position were defined as
 energy-weighted averages of the timings and crystal locations, respectively.
 All the cluster timings were required to be within 10~ns of each other.
 The clusters were required to be outside the
 beam hole region: $\mathrm{max}\{|x|, |y|\}>150$~mm.
 
 The decay vertex position of the $\pi^0 $ along the beam, $z=z_{\rm vtx}^{\pi^0}$, was
 first reconstructed assuming that two of the three clusters were from the $\pi^0$,
 the decay position was along the $z$ axis, and
 the invariant mass, $M_{\gamma_1 \gamma_2}$, was equal to the nominal $\pi^0$ mass.
 Of the three photon combinations, the one with the
 smallest absolute magnitude of two vertex displacement,
 $\Delta z_{\rm vtx}=z_{\rm vtx}^{\pi^0}-z_{\rm vtx}^{K_L}$,
 was selected, where $z_{\rm vtx}^{K_L}$ was calculated
 by assuming that the invariant mass of all the three $\gamma$'s
 equals the nominal mass of $K_L $.
 The four-momenta of the three $\gamma$'s were then reconstructed
 assuming that they are produced at $z_{\rm vtx}^{\pi^0}$. 
 The optimization of selection criteria (cuts) and estimation of acceptances were
 based on the Monte Carlo (MC) simulation using
 the GEANT4 package~\cite{citegeant4_1, citegeant4_2, citegeant4_3}. To 
 reflect the real beam-related activities, the MC events were further
 overlaid with random trigger data taken during physics data collection.
 
To avoid bias, we adopted a blind analysis technique:
 the signal region (SR) had been defined in the two-dimensional space of
 $(z_{\rm vtx}^{\pi^0},M_{\pi^0\gamma})$, and the selection criteria were determined
 using a data set with all the events in the SR removed. 
 To gain the largest possible efficiency for $K_L \to \pi^0 \gamma$ while
 suppressing the background contribution,
 the SR was defined to be $1500~\mathrm{mm}<z_{\rm vtx}^{\pi^0}<3500~\mathrm{mm}$,
 and $490~\mathrm{MeV}<M_{\pi^0\gamma}<520~\mathrm{MeV}$.
 The range of the decay vertex position was
 determined to be more upstream than that of $K_L \to \pi^0 \nu \bar{\nu}$
 analysis ($3000~\mathrm{mm}<z_{\rm vtx}^{\pi^0}<5000~\mathrm{mm}$),
 because the $K_L\to2\pi^0$ decay contributes to the downstream
 region. The
 side band regions were
 used as control regions (CRs).
 The region referred to as CR2,
 defined as $3500~\mathrm{mm}<z_{\rm vtx}^{\pi^0}<6200~\mathrm{mm}$
 and $400~\mathrm{MeV}<M_{\pi^0\gamma}<490~\mathrm{MeV}$, dominated by $K_L\to2\pi^0$ decays, was used
 to calculate the $K_L$ yield. 
 
The shape of each cluster in the $x$-$y$ plane was required
 to be consistent with the shape of electromagnetic shower from a single photon obtained
 by the MC simulation. The MC template of the nominal energy deposits and
 their standard deviations in crystals was prepared as a function of
 the incident angle of $\gamma$ and the observed cluster energy, and was
 used to compute a $\chi^2$ value (shape $\chi^2$).
 We required all the $\gamma$ cluster candidates to satisfy $\chi^2<5.0$.
 These requirements discriminated a hadronic cluster due to neutrons or
 a fusion cluster
 from photon overlaps in close proximity.
 To further remove clusters produced by neutrons,
 a neural network technique (NN)~\cite{TMVA2007} was used to distinguish $\gamma$
 clusters from neutron clusters based on
 the information on two-dimensional cluster shape, relative energies of crystals,
 energy-weighted $xy$-position of the cluster, timings of the observed
 signals in crystals, and the $\gamma$'s incident angle.
 All the $\gamma$'s were required to have a likelihood
 of more than 0.8, which corresponds
 to 90\% efficiency for $\gamma$'s and $\times$~33 reduction for neutrons.

\begin{figure}[]
\begin{center}
{\centering \subfloat[$\Delta z_{\rm vtx}$]{\includegraphics[width=8.3cm,clip]{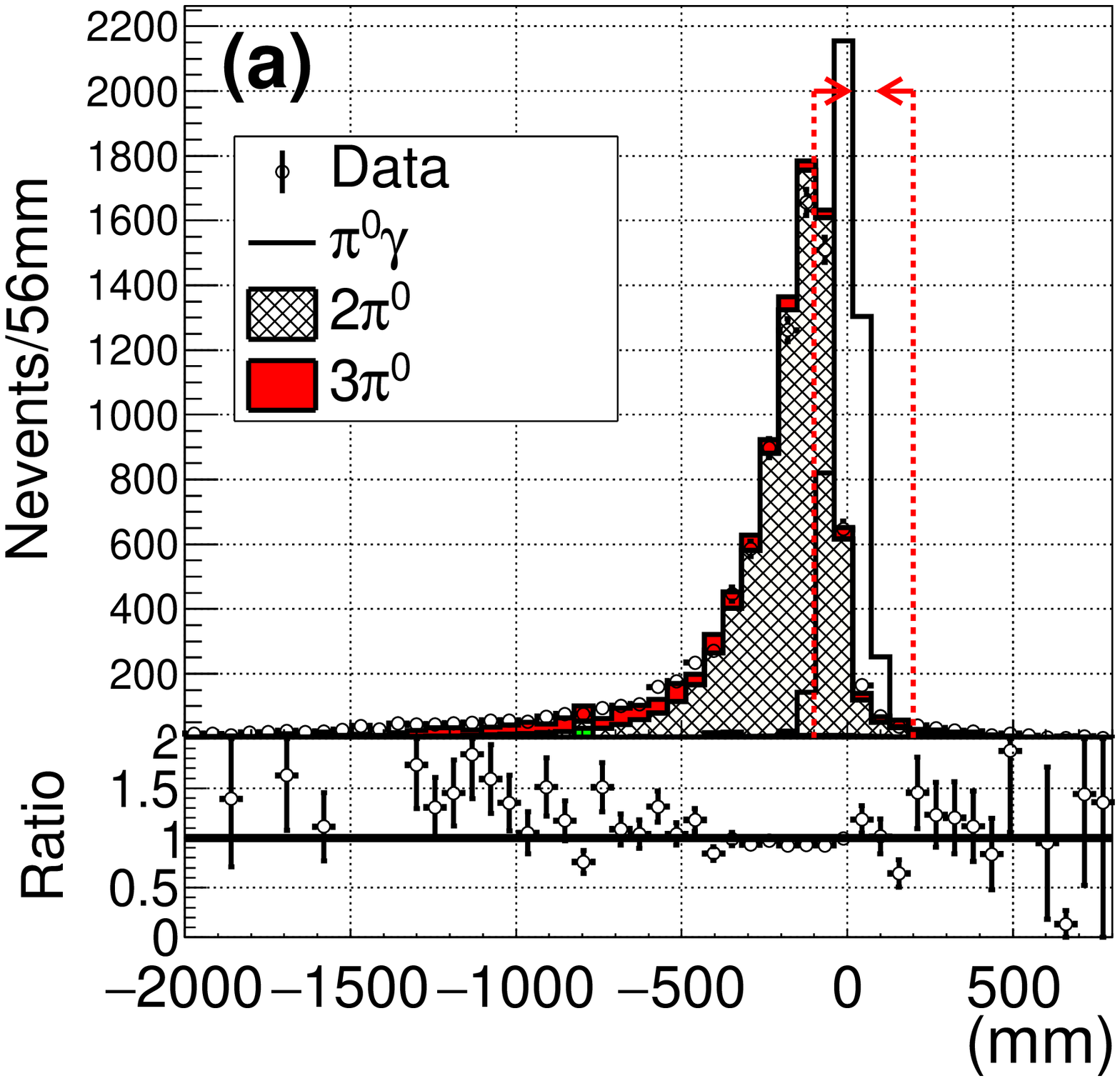} \label{histdelzvtx}} }
{\centering \subfloat[$E_{\gamma}^{\rm min}$]{\includegraphics[width=8.3cm,clip]{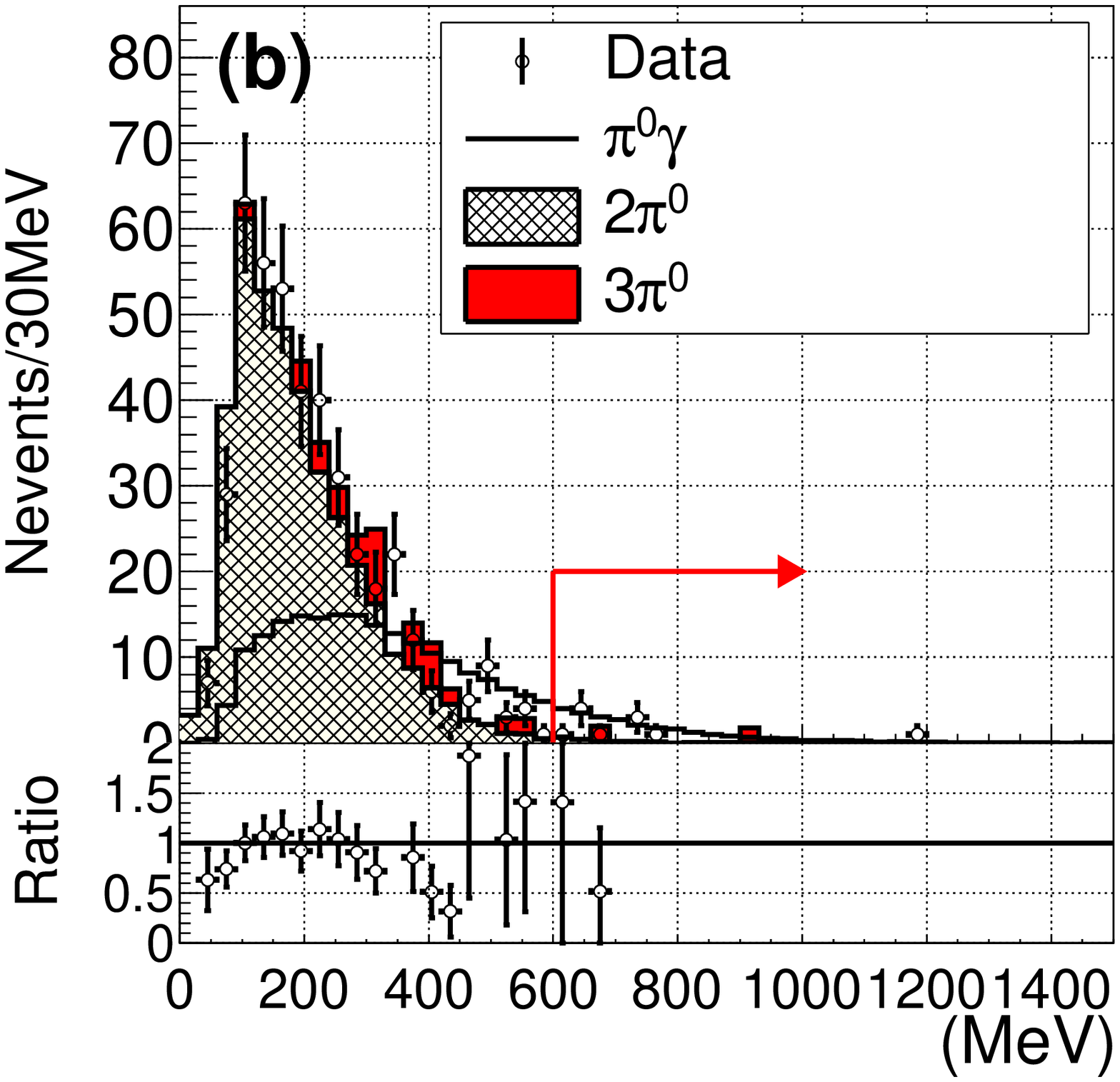} \label{histEmin}} } 
\caption{ (a) Distribution of the two vertex displacement ($\Delta z_{\rm vtx}$) after
 imposing cuts on the shape $\chi^2$, $\gamma/n$ separation with NN, $K_L$ direction, and all the veto counters.
 (b) Distribution of the minimum energy of three $\gamma$'s ($E_{\gamma}^{\rm min}$) after
 imposing the same cuts as (a) and $490~\mathrm{MeV}<M_{\pi^0\gamma}<520~\mathrm{MeV}$.
 Points show data, while histograms of $K_L\to 2\pi^0$ (hatched), $K_L \to 3\pi^0$ (solid),
 and $K_L\to \pi^0\gamma$ (empty), show MC.
 The histogram for the $K_L\to \pi^0\gamma$ decay was
 obtained assuming
 $\mathcal{B}(K_L\to \pi^0\gamma)=\mathcal{B}(K_L\to 2\pi^0)\times 0.01=8.64\times 10^{-6}$~\cite{PDG_paper}.
 Red arrows represent the cuts of $-100~\mathrm{mm}<\Delta z_{\rm vtx}<200~\mathrm{mm}$ (a) and $E_{\gamma}^{\rm min}>600$~MeV (b).
 The bottom regions in both panels show the ratio of data and MC events for
 each histogram bin. For both figures, the events of data in SR were excluded.
} \label{histkine}
\end{center}
\end{figure}

\begin{figure}[]
\begin{center}
{\centering \subfloat[$K_L\to \pi^0 \gamma$]{\includegraphics[width=8.3cm,clip]{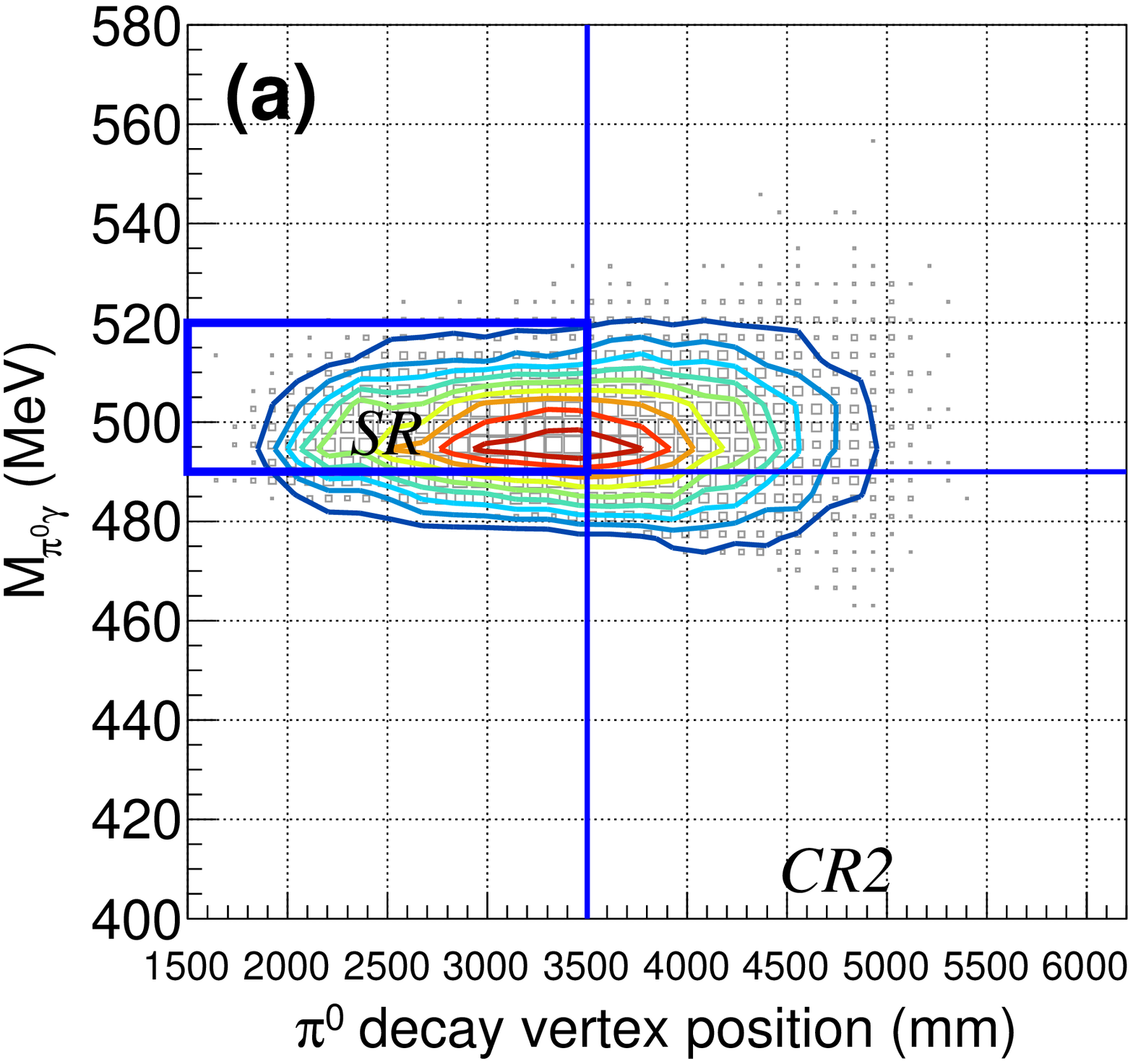} \label{scatterMCa}} }
{\centering \subfloat[Backgrounds]{\includegraphics[width=8.3cm,clip]{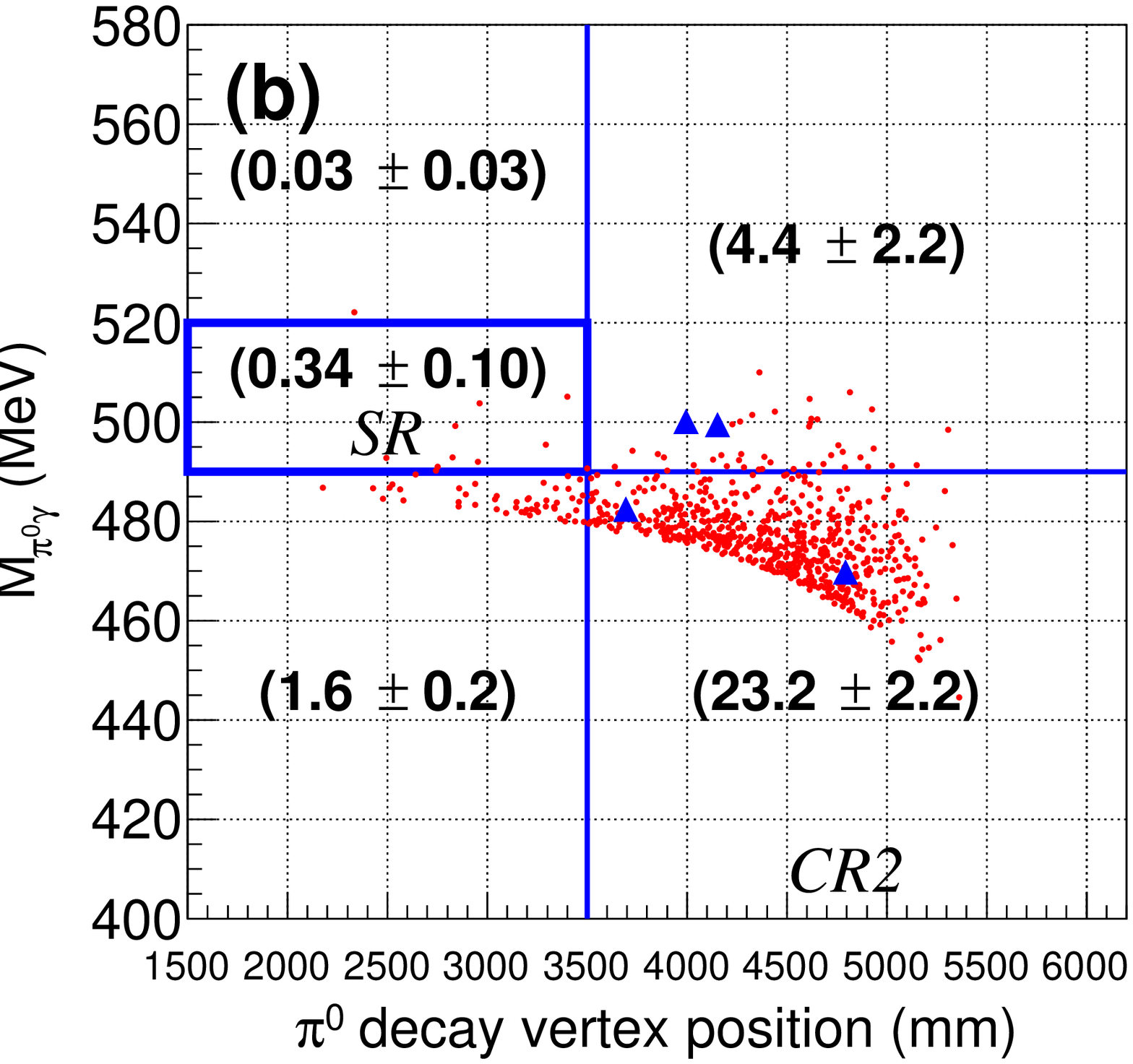} \label{scatterMCb}} } 
\caption{Reconstructed $\pi^0\gamma$ invariant mass ($M_{\pi^0 \gamma}$) versus $\pi^0$ decay point ($z_{\rm vtx}^{\pi^0}$) plot
 by the MC simulation with all cuts imposed on (a) $K_L\to \pi^0 \gamma$
 and (b) background events. The left middle region
 enclosed by thick rectangle is the SR and the bottom right is the CR2.
 The contours in (a) are linear in arbitrary scale.
 The peak and width ($\sigma$ of the Gaussian distribution)
 of the $M_{\pi^0 \gamma}$ distribution
 are 497~MeV and 9.4~MeV, respectively. 
 In (b), the circular and triangular markers represent
 $K_L \to 2 \pi^0$ and $K_L \to 3 \pi^0$ decays, respectively.
 The markers of $K_L \to 2 \pi^0$ and $K_L \to 3 \pi^0$ correspond
 to 0.03 and 1.5 data events, respectively. The numbers
 in parenthesis represent the predicted number of events
 by the MC simulation.
} \label{scatterMC}
\end{center}
\end{figure}

To suppress other $K_L$ decays,
 we required no in-time signals in the veto counters above each threshold.
 In particular, we imposed stringent energy thresholds of $1$~MeV
 in the three barrel counters (FB, MB, and IB) and NCC.
 After imposing all the veto cuts, the $K_L\to 2\pi^0$ decay was
 the largest contribution of all the background sources.
 This decay mode could be a background if a photon with a small energy
 was undetected by the veto counters or two of the four clusters in CSI fused.
 To suppress this contribution, the
 two vertex displacement was required to satisfy
 $-100~\mathrm{mm}<\Delta z_{\rm vtx}<200~\mathrm{mm}$
 as shown in Fig.~\ref{histdelzvtx}. 
 Furthermore, the minimum photon energy of three $\gamma$'s ($E_{\gamma}^{\rm min}$) was
 required to be larger
 than 600~MeV, as shown in Fig.~\ref{histEmin}.
 The $K_L$ momentum was calculated as a sum of momenta of 
 $\gamma_0$, $\gamma_1$, and $\gamma_2$,
 and its transverse momentum and polar angle with respect to the $z$ axis were
 required to be less than 100~MeV$/c$ and $4^{\circ}$,
 respectively ($K_L$ direction cuts). 
 Figures~\ref{scatterMC}a and \ref{scatterMC}b show
 the reconstructed mass ($M_{\pi^0 \gamma}$) versus $\pi^0$ decay vertex ($z_{\rm vtx}^{\pi^0}$)
 plots after imposing the cuts described above for the $K_L\to \pi^0\gamma$ decay
 and other $K_L$ decays generated by MC, respectively.
 A summary of the estimated acceptances for the $K_L\to \pi^0\gamma$, $K_L\to 2\pi^0$, and $K_L\to 3\pi^0$
 decays at each step by the MC samples is shown in Table~\ref{sig_tbl} 
\begin{table*}
\caption{Acceptances of the $K_L\to \pi^0\gamma$, $K_L\to 2\pi^0$, and $K_L\to 3\pi^0$ decays
 at each step of event selection\label{sig_tbl}.}
\begin{ruledtabular}
\scalebox{0.92}[0.92]{
\begin{tabular}{clccc}  
Index&Selection & $K_L\to \pi^0\gamma$ &  $K_L\to 2\pi^0$ & $K_L\to 3\pi^0$\\ \colrule
1& $K_L$ decay \footnote{~~~A probability that
 $K_L$ decay occurs in the SR.} & 9\% & 9\% & 9\% \\
2&Geometry and trigger& $2.2\times 10^{-3}$ & $3.1\times 10^{-4}$ & $1.5\times 10^{-5}$ \\
3&Shape~$\chi^2$ of clusters& $2.0\times 10^{-3}$ & $2.4\times 10^{-4}$ &  $1.1\times 10^{-6}$\\
4& $xy$ position of clusters& $1.9\times 10^{-3}$ & $2.2\times 10^{-4}$ &  $9.7\times 10^{-7}$\\
5&$K_L$ direction & $1.8\times 10^{-3}$ & $2.0\times 10^{-4}$ & $5.8\times 10^{-7}$\\
6&Veto & $7.1\times 10^{-4}$ & $1.9\times 10^{-5}$ & $1.5\times 10^{-8}$\\
7& Separation of $\gamma/n$ with NN & $5.1\times 10^{-4}$ & $1.3\times 10^{-5}$ & $8.2\times 10^{-9}$\\
8&$\Delta z_{vtx}$ & $4.9\times 10^{-4}$ & $4.2\times 10^{-6}$ & $9.1\times 10^{-10}$\\
9&$E_{\gamma}^{\rm min}>$ 300~MeV & $2.4\times 10^{-4}$ & $1.0\times 10^{-6}$ & $4.4\times 10^{-10}$\\
10&$E_{\gamma}^{\rm min}>$ 600~MeV & $5.0\times 10^{-5}$ & $4.1\times 10^{-8}$ & $4.7\times 10^{-11}$\\ \colrule
11&CR2, $E_{\gamma}^{\rm min}>$ 300~MeV & - &  $8.9\times 10^{-7}$ & $1.2\times 10^{-10}$\\
12&SR, $E_{\gamma}^{\rm min}>$ 600~MeV  & $(2.11\pm 0.03)\times 10^{-5}$ & $(5.6\pm 1.8)\times 10^{-10}$ & - \\
\end{tabular}
}
\end{ruledtabular}
\end{table*}. 
 From the indices of 1 to 10, the
 accumulated acceptances are shown, whereas for 11 and
 12, cuts of 1 to 8 are included.

Table~\ref{bg_tabl}
\begin{table}
\caption{Expected numbers of backgrounds in the signal region.
 The upper limits are at 68\% C.L. \label{bg_tabl}}
\begin{ruledtabular}
\begin{tabular}{lc}  
Source & Number of events \\ \colrule
$K_L \to 2\pi^0$ & $0.32 \pm 0.10$  \\
$K_L \to 3\pi^0$ & $<0.5$ \\
$K_L \to 2\gamma $ & $<0.06 $ \\
Neutron & $<0.02$ \\
$K_L \to \pi^0 \gamma \gamma $ & $0.020 \pm 0.002 $ \\
Other $K_L$ decays & $<0.04$ \\ \colrule
Total & $0.34\pm 0.10 ~(<1.0)$  \footnote{
See the main text for this calculation.
The total number of events and its uncertainty in the SR only include background estimates with central values because all of the upper limits came from the limited statistics of MC samples.
 If the contributions of all the sources are assumed,
 the upper limit of the number of backgrounds in the SR is 1.0 at 68\% C.L. 
}   
\end{tabular}
\end{ruledtabular}
\end{table} summarizes the expected number of background events in the SR.
 The contribution of the $K_L\to 2\pi^0$ events
 in the SR was
 estimated to be: \vspace{0mm}
\begin{align}
N^\mathrm{CR2}_{obs}p_{\rm 2\pi^0 }^{\rm CR2}\times \frac{N_{\rm MC}^{\rm SR}}{N_{\rm MC}^{\rm CR2}},
\end{align}
where $N^\mathrm{CR2}_{obs}=528$ is the number of the observed events in the CR2 region
 with an energy threshold for photons of $E_{\gamma}^{\rm min}>300$~MeV,
 $N^{\mathrm{SR}}_{\rm MC}(N^{\mathrm{CR2}}_{\rm MC})$
 is the number of events in the corresponding
 region with an energy threshold of $E_{\gamma}^{\rm min}>$ 600~MeV (300~MeV)
 estimated by the MC simulation, and $p^{\rm CR2}_{2\pi^0}=97\%$
 is the purity of the $K_L \to 2\pi^0$ decay in the CR2.
 
The number of $K_L \to 3\pi^0$ events
 in the SR was estimated based on another control region
 dominated by the $K_L \to 3\pi^0$ decays,
 $ {M_{01}^2+M_{02}^2}>$~$(490$~MeV$)^2$, where $M_{ij}^2={(p_i+p_j)^2}$, and $p_i$ $(i=0,1,2)$
 are the four-vectors of the three $\gamma$'s~\cite{dalitzexp}.
 Using the number of events
 in this region ($N^{\mathrm{CR}}_{obs}=108$) and
 $N^{\mathrm{SR}}_{\rm MC}/N^{\mathrm{CR}}_{\rm MC}<4.3\times 10^{-3}$ at 68\% C.L.,
 we estimated the number of events in the SR to be less than $0.5$ at the 68\% C.L.
 
If the $K_L \to 2\gamma $ decay was coincident with an accidental hit in CSI,
 it could become a background. We generated the MC samples of the $K_L \to 2\gamma $ decay,
 corresponding to 18 times the experimental data, and found that no events satisfied the cuts.
 
Another type of background was the $\pi^0$ production by an
 interaction of beam halo neutrons in NCC, where two photons from
 the $\pi^0$ decay entered CSI with an additional accidental hit.
 We produced a MC sample and confirmed that the $\Delta z_{\rm vtx}$ requirement
 removed all the events in the SR 
 even without imposing various cuts: shape~$\chi^2$, $K_L$ direction, $\gamma/n$ separation with NN, and $E_{\rm min}$.
 We thus set the upper limit to be 0.02 at the 68\% C.L., assuming Poisson distribution.

The $K_L \to \pi^0 \gamma \gamma$ decay mode could be a background to
 $K_L \to \pi^0 \gamma$ if one of the $\gamma$ energies was soft in the laboratory frame.
 However, due to its small branching fraction of
 $\mathcal{B} (K_L \to \pi^0 \gamma \gamma)=1.27\times 10^{-6}$~\cite{PDG_paper},
 its contribution was negligible.
 
We also studied the contributions from $K_L$ decays
 with charged
 particles in the final state.
 The cut on CV hits
 suppressed the
 contributions in the SR to be less than 0.04 events at the 68\% C.L.
 
The total number of background events and its uncertainty in the SR only includes background estimates with central values
 because all of the upper limits came from the
 limited statistics of MC samples. 
 On the other hand, if we conservatively consider the contributions of all the sources,
 the upper limit of the number of backgrounds in the SR was 1.0 at the 68\% C.L.

The branching fraction of the signal was measured using
 the numbers of events in the SR and CR2 as:
\begin{align} 
&\mathcal{B}(K_L\to \pi^0 \gamma) = N^\mathrm{SR}_{obs}\cdot {SES} \nonumber \\
 &= \frac{N^\mathrm{SR}_{obs}}{N^\mathrm{CR2}_{obs}}\cdot \frac{\mathcal{B}(K_L \to 2\pi^0 )\epsilon^{\rm CR2}_{2\pi^0 }+\mathcal{B}(K_L \to 3\pi^0 )\epsilon^{\rm CR2}_{3\pi^0 }}{\epsilon^{\rm SR}_{\pi^0 \gamma} }  , \label{seseq}
\end{align}
where $N^\mathrm{SR}_{obs}$ is the number of observed events
 in the SR, $SES$
 is the single event sensitivity
 of the $K_L \to \pi^0 \gamma$ decay,
 $N^\mathrm{CR2}_{obs}=528$ is the number of the events in
 the CR2 (under the condition of $E_{\gamma}^{\rm min}>300$~MeV),
 $\epsilon^{SR}_{\pi^0 \gamma }=2.1\times 10^{-5}$, $\epsilon^{CR2}_{2\pi^0 }=8.9\times 10^{-7}$,
 and $\epsilon^{CR2}_{3\pi^0 }=1.2\times 10^{-10}$
 are the acceptances obtained by the MC of the $K_L \to \pi^0 \gamma$,
 $K_L \to 2\pi^0$, and $K_L \to 3\pi^0$ decays in each region, respectively,
 and $\mathcal{B}(K_L \to 2\pi^0 )=8.64\times 10^{-4}$ and $\mathcal{B}(K_L \to 3\pi^0 )=19.5\%$
 are the branching fraction of the $K_L \to 2\pi^0$ and $K_L \to 3\pi^0$
 decays, respectively~\cite{PDG_paper}. 
 The obtained $SES$ was
 $(7.1\pm 0.3_{\rm stat.} \pm 1.6_{\rm syst.})\times 10^{-8}$, where the
 first and second uncertainties are statistical and systematic, respectively.

The various sources of uncertainties on $SES$ are summarized in
 Table~\ref{syst_tabl}.
\begin{table}
\caption{Summary of uncertainties in the single event sensitivity.  \label{syst_tabl}}
\begin{ruledtabular}
\begin{tabular}{lc} 
Source & Uncertainty[\%]  \\  \colrule
Offline veto & 17 \\
Kinematic selection & 12 \\
Online veto & 6.4 \\
Online cluster counting & 1.8 \\
Shape $\chi^2$ and $\gamma/n$ separation with NN & 1.5 \\
Geometrical & 1.5 \\
Clustering & 1.0 \\
Reconstruction & 0.3 \\ 
$\mathcal{B}(K_L \to 2\pi^0)$ & 0.6 \\ \colrule
Statistics for normalization & 4.4 \\  \colrule
Total & $4.4_{\rm stat.} \oplus 22_{\rm syst.}$ \\ 
\end{tabular}
\end{ruledtabular}
\end{table}
 As Eq.~\ref{seseq} shows, systematic uncertainties due to common bias between 
 the $K_L \to \pi^0 \gamma$ and $K_L \to 2\pi^0 $ 
 decays cancel ($K_L \to 3\pi^0 $ contribution is only 3\%).
 Thus, we used the events in the CR2 to conservatively evaluate the acceptances
 of offline and online vetoes, common kinematic selection,
 online cluster counting, geometry, and reconstruction.
 Kinematic selections which were not considered to be common were 
 $E_{\gamma}^{\rm min}$, $\Delta z_{\rm vtx}$ and the acceptance being inside of the SR.
 However, the aceptances of these cuts for the $K_L \to 2\pi^0 $ decay are
 much smaller than the $K_L \to \pi^0 \gamma$ decay,
 and the evaluation of the uncertainty of $SES$ by the $K_L \to 2\pi^0 $ decay is also conservative.
 For this reason, we evaluated their uncertainties using the events in the CR2 
 by the loosening $E_{\gamma}^{\rm min}$ threshold down to 300~MeV.
 Shape $\chi^2$, $\gamma/n$ separation with NN, and clustering, were evaluated
 using a different control sample which is dominated by the $K_L \to 3\pi^0$ decays. 
 
The largest contribution
 came from the discrepancies of the offline veto acceptances between data and MC.
 The systematic uncertainty of a given offline cut was calculated using
 a double ratio:
\begin{align}
r = \frac{n_\mathrm{data}/\bar{n}_\mathrm{data}}{n_\mathrm{MC}/\bar{n}_\mathrm{MC}},
\end{align}
where $n_\mathrm{data,MC}$ are the numbers of events after
 imposing all the vetoes, and $\bar{n}_\mathrm{data,MC}$ are
 the corresponding numbers when one of the vetoes was removed.
 The deviation of $r$ from 1 was the systematic
 uncertainty from the offline veto. 
 The quadratic sum of all the vetoes was the
 total systematic uncertainty due to offline veto.
 The second largest effect came from the systematic uncertainty of the
 kinematic selection described before.
 Similarly to the offline veto cuts, we relaxed one of the kinematic cuts and
 compared the double ratio between data and MC.
 This uncertainty was mainly caused by the limited statistics of
 data used for this evaluation.
 The third largest source of the systematic uncertainty was from the online veto.
 This was estimated using data triggered without imposing
 online vetoes while keeping the information on online trigger decision.
 This uncertainty was mainly due
 to the intentionally-loosened
 offline veto energy threshold of CV
 set to minimize the acceptance loss from the accidental hits.
 As a result, the online threshold was close to that of the offline threshold.
 Uncertainties from online cluster counting, cluster-shape discrimination,
 geometrical acceptance,
 clustering, and reconstruction were smaller than those from the three aforementioned sources.
 The uncertainty of the branching fraction of $K_L \to 2\pi^0$ was taken from the PDG value~\cite{PDG_paper}.
 The total statistical and systematic uncertainties were 4.4\% and 22\%, respectively.

\begin{figure}[h]
\begin{center}
{\centering { \includegraphics[width=8.3cm,clip]{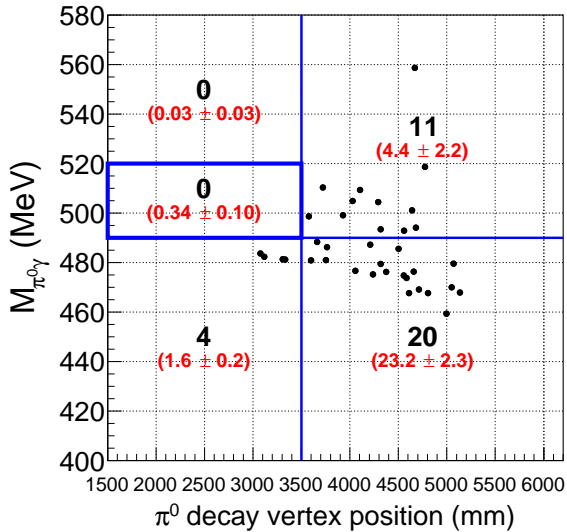}} }
\caption{Reconstructed $\pi^0\gamma$ invariant
 mass ($M_{\pi^0 \gamma}$) versus $\pi^0$ decay point ($z_{\rm vtx}^{\pi^0}$) plot with all cuts imposed on
 data.
 Values are the numbers of the observed events and lower ones (in parenthesis)
 are predictions by the MC simulation.} \label{scatterDT}
\end{center}
\end{figure}

After determining the cuts described above, we unmasked the SR
 and observed no candidate events, as shown in Fig.~\ref{scatterDT}. 
 The discrepancy between the number of observed events and the MC simulation
 in the upper right region could be explained by the limited
 statistics of the simulated $K_L\to3\pi^0$ decay sample.
 In fact, when we loosened the cut on
 the minimum photon energy ($E_{\gamma}^{\rm min}$) from 600~MeV to 300~MeV,
 the contribution
 from the $K_L\to3\pi^0$ and $K_L\to2\pi^0$ decays by the MC simulation
 increased to $41\pm 7$ and $47\pm 1$, respectively, in which uncertainties
 were statistical only,
 and the sum of them was consistent with the observation of 96 events.
 Taking into account the systematic uncertainty of $SES$~\cite{Cousins},
 the upper limit was set to $\mathcal{B}(K_L \to \pi^0 \gamma)<1.7 \times 10^{-7}$ at the 90\% C.L.
 This is the first experimental upper limit set on the $K_L \to \pi^0 \gamma$ decay.

In conclusion, we searched for the $K_L \to \pi^0 \gamma$ decay, which is forbidden by
 Lorentz invariance and gauge invariance, for the first time.
 From the data collected between 2016 and
 2018, we observed no candidate events in the signal region.
 The first upper limit of the branching fraction of the $K_L \to \pi^0 \gamma$ decay is
 $1.7\times 10^{-7}$ at the 90\% confidence level.

\section*{Acknowledgement}
We would like to express our gratitude to all members of
 the J-PARC Accelerator and Hadron Experimental Facility groups for their support.
 We also thank the KEK Computing Research Center for KEKCC
 and the National Institute of Information for SINET4.
 This work was supported by the Ministry of Education,
 Culture, Sports, Science, and Technology (MEXT) of
 Japan and the Japan Society for the Promotion of
 Science (JSPS) under the MEXT KAKENHI Grant No.~JP18071006,
 the JSPS KAKENHI Grants No.~JP23224007,
 and No.~JP16H06343, by the research
 fellowship program
 for postdoctoral scientists No.~17J02178, and
 through the Japan-U.S. Cooperative Research
 Program in High Energy Physics; the U.S. Department of Energy,
 Office of Science, Office of High Energy Physics,
 under Awards No.~DE-SC0006497, No.~DE-SC0007859, and
 No.~DE-SC0009798; the Ministry of Education and the Ministry
 of Science and Technology in Taiwan under Grants
 No.~104-2112-M-002-021, 105-2112-M-002-013 and
 106-2112-M-002-016; and the National Research Foundation of Korea
 (2017R1A2B2011334 and 2019R1A2C1084552).


\begin{thebibliography}{99}

\bibitem{decayofheavymesons} S.~Oneda, S.~Sasaki, and S.~Ozaki, Prog.\ Theo.\ Phys.\ {\bf 5}, 165~(1950).
\bibitem{HighEnergytest} S.~Coleman, and S.~L.~Glashow, Phys.\ Rev.\ D\ {\bf 59},\ 116008\ (1999).
\bibitem{lv_spect} Y.~Michimura, N.~Matsumoto, N.~Ohmae, W.~Kokuyama, Y.~Aso, M.~Ando, and K.~Tsubono, Phys.\ Rev.\ Lett.\ {\bf 110},\ 200401\ (2013).
\bibitem{lv_spect2} M.~E.~Tobar, P.~Wolf, S.~Bize, G.~Santarelli, and V.~Flambaum, Phys.\ Rev.\ D\ {\bf 81},\ 022003\ (2010).
\bibitem{noncummutivity} B.~Meli\'{c}, K.~P.~Kumeri\u{c}ki, and J.~Trampeti\'{c}, Phys.\ Rev.\ D\ {\bf 72},\ 057502 (2005).
\bibitem{e949results} A.~V.~Artamonov {\it et al.}, Phys.\ Lett.\ B\ {\bf 623},\ 192\ (2005).
\bibitem{citebeamline} T. Shimogawa, Nucl.\ Instrum.\ Methods\ A\ {\bf 623}, 585 (2010).
\bibitem{KLpi0nunutheory} A.~J.~Buras {\it et al.}, J.\ High\ Energy\ Phys.\ {\bf 11}\ 033\ (2015).
\bibitem{citeCSI} E.~Iwai {\it et al.}, Nucl.\ Instrum.\ Methods\ A\ {\bf 786}\ 135\ (2015).
\bibitem{citeCSI2} T. Masuda {\it et al.}, Nucl.\ Instrum.\ Methods\ A\ {\bf 746},\ 11\ (2014).
\bibitem{IBpaper} R.~Murayama, M.~Togawa {\it et al.}, Nucl.\ Instrum.\ Methods\ A\ {\bf 953},\ 163255\ (2020).
\bibitem{citeCV} D.~Naito {\it et al.}, Prog.\ Theor.\ Exp.\ Phys.,\ {\bf 2016},\ 023C01\ (2016).
\bibitem{cite_BHPV} Y.~Maeda {\it et al}., Prog.\ Theor.\ Exp.\ Phys.,\ {\bf 2015},\ 063H01\ (2015).
\bibitem{KOTOdet} J.~K.~Ahn {\it et al.}, Prog.\ Theor.\ Exp.\ Phys.,\ {\bf 2017},\ 021C01\ (2017).
\bibitem{KOTO2015} J.~K.~Ahn {\it et al.}, Phys.\ Rev.\ Lett.\ {\bf 122},\ 021802\ (2019). 
\bibitem{JayDAQpaper} C.~Lin {\it et al.}, J.\ Phys.\ Conf.\ {\bf 1526},\ 012034\ (2020).
\bibitem{cite_masuda} T. Masuda {\it et al.}, Prog.\ Theor.\ Exp.\ Phys.,\ {\bf 2016},\ 013C03\ (2016).
\bibitem{citegeant4_1} S.~Agostinelli {\it et al.}, Nucl.~Instrum.~Methods~A~{\bf 506}, 250 (2003).
\bibitem{citegeant4_2} J. Allison {\it et al.}, IEEE\ Trans.\ Nucl.\ Sci.\ {\bf 53},\ 270\ (2006).
\bibitem{citegeant4_3} J. Allison {\it et al.}, Nucl.\ Instrum.\ Methods\ A\ {\bf 835},\ 186\ (2016).
\bibitem{TMVA2007} A.~Hoecker {\it et al.}, PoS\ A\ CAT\ {\bf 040} [physics/0703039] (2007).
\bibitem{PDG_paper} M.~Tanabashi {\it et al.}, (Particle Data Group), Phys.\ Rev.\ D\ {\bf 98},\ 030001\ (2018).
\bibitem{dalitzexp} For the $K_L\to\pi^0 \gamma $ decay,
 $ {M_{01}^2+M_{02}^2}=M_{K_L}^2-M_{\pi^0}^2=(479~\mathrm{MeV})^2$. This value
 becomes smaller when the $K_L\to2\pi^0$ decay makes
 a three cluster event with missing either $\gamma$,
 whereas this value can become larger when the $K_L\to3\pi^0$ decay makes a three cluster event
 with fusion clusters.
\bibitem{Cousins} R.~D.~Cousins, and V.~L.~Highland, Nucl.\ Instrum.\ Methods\ A\ {\bf 320},\ 331\ (1992).
\end{thebibliography}
\end{document}